\documentclass[preprint,12pt]{elsarticle}

\usepackage{amssymb}

\begin{document}

\begin{frontmatter}



\title{Universal power laws in the threshold network model: \\
       A theoretical analysis based on extreme value theory}


\author{A. Fujihara\corref{cor1}}
  \ead{afujihara@kwansei.ac.jp}
  \cortext[cor1]{Corresponding author}
  \address{Graduate School of Science and Technology, Kwansei Gakuin 
           University, \\
           2-1 Gakuen Sanda, Hyogo 669-1337, Japan}

\author{M. Uchida}
  \address{Network Design Research Center, Kyushu Institute of Technology, \\
      2-2-3 Uchisaiwaicho Chiyoda-ku, Tokyo 100-0011, Japan}

\author{H. Miwa}
  \address{Graduate School of Science and Technology, Kwansei Gakuin 
           University, \\
           2-1 Gakuen Sanda, Hyogo 669-1337, Japan}

\begin{abstract}
 We theoretically and numerically investigated the threshold network model 
with a generic weight function where there were a large number of nodes 
and a high threshold. 
 Our analysis was based on extreme value theory, which gave us a 
theoretical understanding of the distribution of independent and identically
distributed random variables within a sufficiently high range.
 Specifically, the distribution could be generally expressed by 
a generalized Pareto distribution, which enabled us to formulate 
the generic weight distribution function.
 By using the theorem, we obtained the exact expressions of 
degree distribution and clustering coefficient which behaved as 
universal power laws within certain ranges of degrees. 
 We also compared the theoretical predictions with numerical results 
and found that they were extremely consistent. 
\end{abstract}

\begin{keyword}
  Complex Networks \sep Threshold Network Model 
  \sep Extreme Value Theory \sep Power laws 
  \PACS 89.75.Fb \sep 89.75.Da
\end{keyword}

\end{frontmatter}


\section{Introduction}
\label{sec:intro}
 Many researchers during the last decade have enthusiastically 
investigated
the structure of real networks that have emerged from various disciplines, 
such as information science, sociology, and biology.
 As a result of these investigations, some have found two features inherent 
in their networks, i.e., \textit{small-world} \cite{WS1998} and 
\textit{scale-free} \cite{BA1999},
which were vastly different from those of the regular and random networks
already known. 
 The term ``small-world" means that the clustering coefficient 
remains large and that there are very short paths between every two nodes, 
and ``scale-free" means that the degree distribution has a long tail, 
which usually obeys a power law. 
 All networks having both (or either) of these are often called 
complex networks. Many physicists have tried to theoretically understand 
their features by using mathematical models. 

 The threshold network model is a non-growing network model that explains
the properties of complex networks. 
 Each node in this model has an intrinsic weight that is randomly distributed 
according to a certain probability distribution function.
 Every pair of nodes satisfying the condition that the sum of their 
weights is over a certain given threshold is linked. 
 When the weights are generated from various types of restricted 
distribution (specifically, exponential or power-law type distributions),
both the degrees of nodes and the degree-wise clustering coefficient obey
universal power laws at the high tails of 
the degrees \cite{CCDM2002,MMK2004,MMK2005}.
 We investigated the scale-free feature for both the degrees and 
clustering coefficient in a wider variety of weight distribution functions
using extreme value theory\cite{EKM1997,Coles2001,HF2006}. 
 Extreme value theory provides an explanation of the distribution of
independent and identically distributed (iid) random variables 
within a sufficiently high range. 
According to this theory, the distribution is universally expressed 
with a generalized Pareto distribution (GPD). 
 This gives the expression of a generic weight distribution function, 
which finally results in universal power laws, which is the main 
contribution of this paper. 

 This paper is organized as follows. 
We briefly explain the threshold network model in Sec.~\ref{sec:tnm},
then give the formulas for the distributions of degrees and clustering 
coefficient. 
 In Sec.~\ref{sec:evt}, we briefly review extreme value theory.
We present the theoretical results of universal power laws in 
Sec.~\ref{sec:main_result} and compare the theory with numerical results 
in Sec.~\ref{sec:simulation}.
 We finally conclude with a summary and a discussion in Sec.~\ref{sec:SD}.

\section{\label{sec:tnm}Threshold Network Model}
 First, we will explain the threshold network 
model \cite{CCDM2002,MMK2004,MMK2005} with $n (> 1)$ nodes. 
 We assign a weight, $w_{i} \ (1 \le i \le n)$, to each node
by using an iid random variable from a certain weight distribution 
function, $f(w)$. 
 When the sum of two weights $w_i, w_j$\ ($i \neq j$) in every pair of nodes 
is more than or equal to a given threshold, $\theta$, i.e.,
\begin{equation}
w_i + w_j \ge \theta. 
\end{equation}
 an edge is formed between them. Going through this procedure, 
we can finally obtain a threshold network. 

 Let
\begin{equation}
F(w) = \int_{-\infty}^{w}f(w')dw' \label{eq:weight_function}
\end{equation}
be a weight function. 
 Here, we assumed that the threshold network had a large number of nodes, 
$n$. From the definition of the threshold network, nodes with the same weight 
values have an equal number of degrees, and the degrees of nodes monotonically 
increase as their weight values increase.
 Due to these features, the degree of the threshold model is specifically 
described in the continuous limit of $n$ as 
\begin{equation}
k = n\int_{\theta - w}^{\infty}f(w')dw' = n\{1-F(\theta - w)\},
\label{eq:degree1}
\end{equation}
 where $0 \le k \le n$. This equality gives a one-to-one correspondence 
between the degree, $k$, and the weight, $w$. 
 Using Eq.~(\ref{eq:degree1}), the degree distribution, 
$p(k)$ ($0 \le k \le n$), naturally follows 
\begin{equation}
p(k) \equiv f(w)\frac{dw}{dk} = 
\frac{f\biggl(\theta - F^{-1}\biggl(1-\frac{k}{n}\biggr)\biggr)}{nf\biggl(F^{-1}\biggr(1-\frac{k}{n}\biggr)\biggr)}.
\label{eq:degree_dist}
\end{equation}
 The degree-wise clustering coefficient, $C(k)$, means that the clustering 
coefficient of nodes with degree $k$ also leads to
\newpage
\begin{eqnarray}
C(k) \equiv & \frac{1}{k(k-1)} \biggl\{ \int_{w}^{\infty} dw' n f(w') \int_{\theta - w}^{\infty} dw'' n f(w'') \nonumber \\
            & + \int_{\theta - w}^{w} dw' n f(w') \int_{\theta - w'}^{\infty} dw'' n f(w'') \biggl\} \nonumber \\
           =&\frac{n^2}{k(k-1)}\biggl\{-1 + 2\frac{k}{n} + \biggl(1-\frac{k}{n}\biggr)F\biggl(\theta - F^{-1}\biggl(1-\frac{k}{n}\biggr)\biggr) \nonumber\\ 
            &- \int_{n[1-F(\theta - F^{-1}(1-\frac{k}{n}))]}^{k}\biggl(1-\frac{k'}{n}\biggr)p(k')dk' \biggr\}.  
\label{eq:cluster_coeff}
\end{eqnarray}

\section{Extreme Value Theory}
\label{sec:evt}
 Next, let us review the essence of extreme value theory.
This theory mathematically treats the statistical behavior of the 
\textit{maximum} 
\begin{equation}
M_n = \max\{X_1, \cdots, X_n\},
\end{equation}
where $X_i$ ($1 \le i \le n$) is an ensemble of iid random variables 
given by a common distribution function, $F(x)$. 
 In a linearly renormalized maximum, there exists the following limit 
distribution
\begin{equation}
G(z) = \exp \left\{ -\left[ 1 + \xi \left( \frac{z-\mu}{\sigma} \right) \right]^{-1/\xi} \right\}, \label{eq:gev_fam_dist}
\end{equation}
 where $z$ is defined as $ \{ z : 1 + \xi ( z - \mu ) / \sigma > 0 \} $, 
$-\infty < \mu < \infty$ is a location parameter, $\sigma > 0$ is a 
scale parameter, and $-\infty < \xi < \infty$ is a shape parameter. 
 This is called a distribution of generalized extreme values (GEVs).
 Note that the GEV distribution includes three statistical distributions: 
Fr\'echet ($\xi > 0$), Gumbel ($\xi = 0$), and (reversed, meaning 
$x \to -x$) Weibull ($\xi < 0$).
 As listed in Table \ref{tab:table1}, the maxima of various types of 
distribution function converge to a GEV distribution (for more details, 
see pp. $153$--$157$ in Embrechts \textit{et al.} \cite{EKM1997}). 
%
\begin{table*}
\caption{\label{tab:table1} Relation between types of distribution 
function $F(x)$ and shape parameter $\xi$. }
\begin{tabular}{|c|c|c|}
\hline
Name & Distribution Function $F(x)$ & Shape Parameter $\xi$ \\
\hline
     & & \\
Beta & $\frac{\Gamma(a+b)}{\Gamma(a) \Gamma(b+1)} \int_{0}^{x} x'^{a-1} (1-x')^{b-1} dx',$ & $-1/b$ \\
     & $0 < x < 1,\ \ a, b > 0$. & \\
     & & \\
Uniform & $x,\ \ \ \ \ 0 < x < 1.$ & $-1$ \\
     & & \\
Exponential & $1 - \exp(-\lambda x),\ \ \ \ \ x > 0,\ \ \lambda > 0.$ & $0$ \\
     & & \\
Normal & $\int_{-\infty}^{x} \frac{1}{s \sqrt{2 \pi}} \exp\{-(x'-m)^{2}/ 2 s^{2}\} dx',$ & $0$ \\
       & $-\infty < x < \infty,\ \ -\infty < m < \infty,\ \ s > 0$. & \\
     & & \\
Lognormal & $\int_{0}^{x} \frac{1}{S \sqrt{2 \pi}} \exp\{ -(\ln x'-M)^{2}/2S^{2} \} dx'/x',$ & $0$ \\
          & $x > 0,\ \ -\infty < M < \infty,\ \ S > 0$. & \\
     & & \\
Pareto & $1 - x^{-\alpha},\ \ \ \ \ x > 1,\ \ \alpha > 0.$ & $1/\alpha$ \\
     & & \\
\hline
\end{tabular}
\end{table*}

 It is well known in extreme value theory that the following theorem 
is satisfied using $G(x)$.
\newtheorem{thm}{Theorem}
\begin{thm}
 Assume that the sample maxima of $F(x)$ converge to $G(x)$.
 Then, for sufficiently large $u$, a distribution function $H$ of $Y = X - u$, 
conditional on $X > u$, yields 
\begin{eqnarray}
H_u(y) &\equiv& \textrm{Pr}\{ X \le y+u\ |\ X>u \} \nonumber \\
            &=& \frac{F(u+y) - F(u)}{1-F(u)} \sim 1 - \left( 1+\frac{\xi y}{\tilde{\sigma}} \right)^{-1/\xi},
\label{eq:huy} 
\end{eqnarray}
on  $ \{ z : 1 + \xi y / \tilde{\sigma} > 0 \} $, where $\tilde{\sigma} = \sigma + \xi(u-\mu)$. 
\end{thm}
 The last equality in Eq.~(\ref{eq:huy}) is called a generalized Pareto 
distribution (GPD).
 For details of the proof, see pp. $76$--$77$ in Coles \cite{Coles2001} or
pp. $10$--$11$ in de Haan and Ferreira \cite{HF2006}. 

 We rearrange Eq.~(\ref{eq:huy}) with respect to $F(x)$, where $x=y+u$. 
It then follows that 
\begin{eqnarray}
F(x) &=   & (1-F(u)) H_u(x-u) + F(u), \nonumber \\
     &\sim& 1 - (1-F(u)) \left( 1+ \frac{\xi(x-u)}{\tilde{\sigma}} \right)^{-1/\xi}, \label{eq:evt_fx}
\end{eqnarray}
for $x \ge u$. 
 Here, assume $F(x)$ in Eq.~(\ref{eq:evt_fx}) is a weight distribution 
function of the threshold network in Eq.~(\ref{eq:weight_function}). 
 Then, Eq.~(\ref{eq:evt_fx}) represents a generic weight distribution 
function under the conditions where the number of nodes $n$ and the lower 
bound, $u$, in Eq.~(\ref{eq:evt_fx}) are sufficiently large.

\section{Main Results}
\label{sec:main_result}

\subsection{Degree Distribution}
\label{subsubsec:ntm_degree}
 To calculate the degree distribution in Eq.~(\ref{eq:degree_dist})
with the generic weight function in Eq.~(\ref{eq:evt_fx}), 
we first need to derive $f(x)$ and $F^{-1}(1-k/n)$. 
 The former can easily be calculated by differentiating Eq.~(\ref{eq:evt_fx}) : 
\begin{equation}
f(x) \equiv F'(x) = \frac{1-F(u)}{\tilde{\sigma}} \left( 1 + \frac{\xi (x-u)}{\tilde{\sigma}} \right)^{-\frac{1}{\xi}-1}, \label{eq:fprimex}
\end{equation}
for $x \ge u$. 
 The latter can also be calculated by using Eqs.~(\ref{eq:degree1}) 
and (\ref{eq:evt_fx}).
 Rearranging Eq.~(\ref{eq:degree1}) with respect to $\theta - w$, we have 
\begin{equation}
F^{-1}\left( 1 - \frac{k}{n} \right) = \theta - w.
\label{eq:F_inv0}
\end{equation}
 Here, substituting Eq.~(\ref{eq:evt_fx}) with $x=\theta-w \ (\ge u)$ into
Eq.~(\ref{eq:degree1}), it follows that
\begin{eqnarray}
\frac{k}{n} &=& 1-F(\theta-w) \nonumber \\
            &=& (1-F(u))\left( 1+\frac{\xi}{\tilde{\sigma}}(\theta-w-u)
                \right)^{-\frac{1}{\xi}}, \label{eq:app_evt_Fx} 
\end{eqnarray}
where $\theta - w \ge u$. 
 Rearranging the last equality in Eq.~(\ref{eq:app_evt_Fx}) with respect to
$\theta - w$, and then substituting this resulting equation into 
the right-hand side of Eq.~(\ref{eq:F_inv0}), it follows that
\begin{eqnarray}
F^{-1}\left( 1-\frac{k}{n} \right) = \frac{\tilde{\sigma}}{\xi} \left\{ \left( \frac{k/n}{1-F(u)} \right)^{-\xi} - 1 \right\} + u. \label{eq:F_inv}
\end{eqnarray}
 Therefore, by using Eqs.~(\ref{eq:fprimex}) and (\ref{eq:F_inv}), 
the degree distribution of the generic weight function is finally given
by
\begin{equation}
p(k) = \frac{1}{n} \left[ \left\{ 2 + (\theta - 2u)\frac{\xi}{\tilde{\sigma}} \right\} \left( \frac{k/n}{1-F(u)} \right)^{\xi} - 1 \right]^{-\frac{1}{\xi}-1}. \label{eq:p_k_ngtm_exact} 
\end{equation}
 As we can see from Eq.~(\ref{eq:p_k_ngtm_exact}), the degree distribution
obeys a power law, $p(k) \sim k^{-(1+\xi)}$ when $\xi \neq 0$.
 For $\xi = 0$, the distribution becomes
\begin{eqnarray}
p(k) &=  & d(k) \exp\{-c(k)\}, \nonumber \\
     &\to& \frac{ n (1-F(u))^2 \exp\left\{ -\frac{\theta - 2u}{\tilde{\sigma}}\right\}}{k^{2}} \sim k^{-2}, \label{eq:pk_xi_zero}
\end{eqnarray}
for $\xi \to 0$, where
\begin{eqnarray}
c(k) &=& \frac{1}{\xi} \ln \left[ \left\{ 2 + (\theta - 2u)\frac{\xi}{\tilde{\sigma}} \right\} \left( \frac{k/n}{1-F(u)} \right)^{\xi} - 1 \right], \nonumber \\
d(k) &=& \frac{1}{n} \left[ \left\{ 2 + (\theta - 2u)\frac{\xi}{\tilde{\sigma}} \right\} \left( \frac{k/n}{1-F(u)} \right)^{\xi} - 1 \right]^{-1}. \nonumber 
\end{eqnarray}
 To summarize, we analytically obtain the universal power law for the 
range of large degree, 
\begin{eqnarray}
p(k) &\sim& k^{-(1+\xi)} \hspace{5mm} (\xi \neq 0), \label{eq:power-law1a} \\
     &\sim& k^{-2} \hspace{11mm} (\xi = 0). \label{eq:power-law1b}
\end{eqnarray}

 Here, note that Eqs.~(\ref{eq:fprimex}) and ~(\ref{eq:app_evt_Fx}) that
we used to obtain the power law are satisfied for $x \ge u$ and 
$\theta - w \ge u$, respectively. Therefore, the applicable range of weight
in the power law is given by
\begin{equation}
u \le w \le \theta - u. \label{eq:w_conditions}
\end{equation}
 Equivalently, that of degrees is also obtained by using Eq.~(\ref{eq:degree1}) 
\begin{equation}
n \{ 1 - F(\theta - u) \} \le k \le n \{ 1 - F(u) \}. \label{eq:corr_range_k}
\end{equation}
 These inequalities indicate the applicable range of degree-distribution 
scales with the number of nodes, $n$.
 To keep the inequalities in Eq.~(\ref{eq:w_conditions}) unbroken, 
threshold $\theta$ must satisfy $\theta \ge 2u$, where $u$ is sufficiently 
large. This indicates that we have to implicitly assume a high threshold to
universally observe the power laws.

\subsection{Clustering Coefficient}
\label{subsubsec:ntm_cluster}
 We next calculate the clustering coefficient in
Eq.~(\ref{eq:cluster_coeff}) with the generic weight function in
Eq.~(\ref{eq:evt_fx}). This automatically means that 
the clustering coefficient is also restricted in the range of
Eqs.~(\ref{eq:corr_range_k}) and ~(\ref{eq:w_conditions}), 
resulting in a high threshold. 
 First, we need to check whether the lower bound of the integral range 
in Eq.~(\ref{eq:cluster_coeff}) is within the range in 
Eq.~(\ref{eq:corr_range_k}). 
 Using Eqs.~(\ref{eq:F_inv0}) and ~(\ref{eq:w_conditions}), 
it follows that 
\begin{eqnarray}
n\left[ 1 - F\left( \theta - F^{-1} \left( 1 - \frac{k}{n} \right)
    \right) \right] &=  & n\{ 1 - F(w) \} \nonumber \\
                    &\ge& n\{ 1 - F(\theta - u) \}, \nonumber
\end{eqnarray}
 This means that the lower bound is inside the domain in
Eq.~(\ref{eq:corr_range_k}), and therefore the integral can successfully
be calculated without violating the requirements of extreme value theory. 
 Combining Eqs.~(\ref{eq:evt_fx}) and ~(\ref{eq:F_inv}), 
it follows that
\begin{eqnarray}
F\left( \theta - F^{-1}\left( 1 - \frac{k}{n} \right) \right) = 1-(1-F(u)) D, \label{eq:FD}
\end{eqnarray}
where 
\begin{eqnarray}
D = \left\{ 2 + \frac{\xi}{\tilde{\sigma}}(\theta-2u) - \left( \frac{k/n}{1-F(u)} \right)^{-\xi} \right\}^{-\frac{1}{\xi}}. \label{eq:D}
\end{eqnarray}
 Therefore, the clustering coefficient is given by
\begin{eqnarray}
C(k) &=   & \frac{n^2}{k(k-1)} \left\{ \frac{k}{n} -  \left( 1 - \frac{k}{n} \right) (1-F(u)) D \right . \nonumber \\
     &    & - \left. \int_{n (1 - F(u)) D}^{k} \left( 1 - \frac{k'}{n} \right) p(k') dk' \right\}, \label{eq:ck_integral_form} \\
     &\sim& k^{-1}, \nonumber
\end{eqnarray}
for large $k$, where $p(k)$ is the same as in Eq.~(\ref{eq:p_k_ngtm_exact}). 
 When $\xi = 0$, Eq.~(\ref{eq:FD}) with Eq.~(\ref{eq:D}) can be explicitly
calculated as
\begin{eqnarray}
F\left( \theta - F^{-1}\left( 1 - \frac{k}{n} \right) \right) \nonumber \\
= 1 - \frac{(1-F(u))^2 n \exp\left\{ -\frac{\theta - 2u}{\tilde{\sigma}} \right\} }{k}. \label{eq:xi_to_zero}
\end{eqnarray}
 When $\xi = 0$, therefore, 
\begin{eqnarray}
C(k) &=   & \frac{n^2}{k(k-1)} (1 - F(u))^2 \exp\left( \frac{2 u - \theta}{\tilde{\sigma}} \right) \nonumber \\
     &    & \times \left( 1 + \frac{\theta - 2u}{\tilde{\sigma}} + 2 \log \frac{k}{n(1-F(u))} \right), \nonumber \\
     &\sim& k^{-2}. \label{eq:Ck_xi_zero}
\end{eqnarray}
 To sum up, the clustering coefficient also obeys a universal power
law expressed by
\begin{eqnarray}
C(k) &\sim& k^{-1} \hspace{5mm} (\xi \neq 0), \label{eq:powerlaw2a} \\
     &\sim& k^{-2} \hspace{5mm} (\xi = 0). \label{eq:powerlaw2b}
\end{eqnarray}

 Finally, we should mention $C(k) = 1$ is always satisfied in the range of
degree $k \le n \{ 1 - F(\theta / 2) \}$ or equivalently when the weight
$w \le \theta / 2$. 
 Since $u \le \theta / 2 \le \theta - u$, it immediately follows that 
$\theta \ge 2 u$. Therefore, 
\begin{equation}
n \{ 1 - F(\theta - u) \} \le n \{ 1 - F(\theta / 2) \} \le n \{ 1 - F(u) \}. \label{eq:ck_ineq}
\end{equation}
 These results mean that $C(k) = 1$ for 
$n \{ 1 - F(\theta - u) \} \le k \le n \{ 1 - F(\theta / 2) \}$, and 
more importantly, the range of the clustering coefficients in 
Eqs.~(\ref{eq:powerlaw2a}) and ~(\ref{eq:powerlaw2b}) is modified by
\begin{equation}
n \{ 1 - F(\theta / 2) \} \le k \le n \{ 1 - F(u) \}. \label{eq:true_ck_range}
\end{equation}
 As well as the case of the degree distribution, the applicable range of 
the clustering coefficients is also scaled with the number of nodes, $n$.

\section{Numerical results}
\label{sec:simulation}
\begin{figure*}[tb]
\resizebox{70mm}{!}{\includegraphics{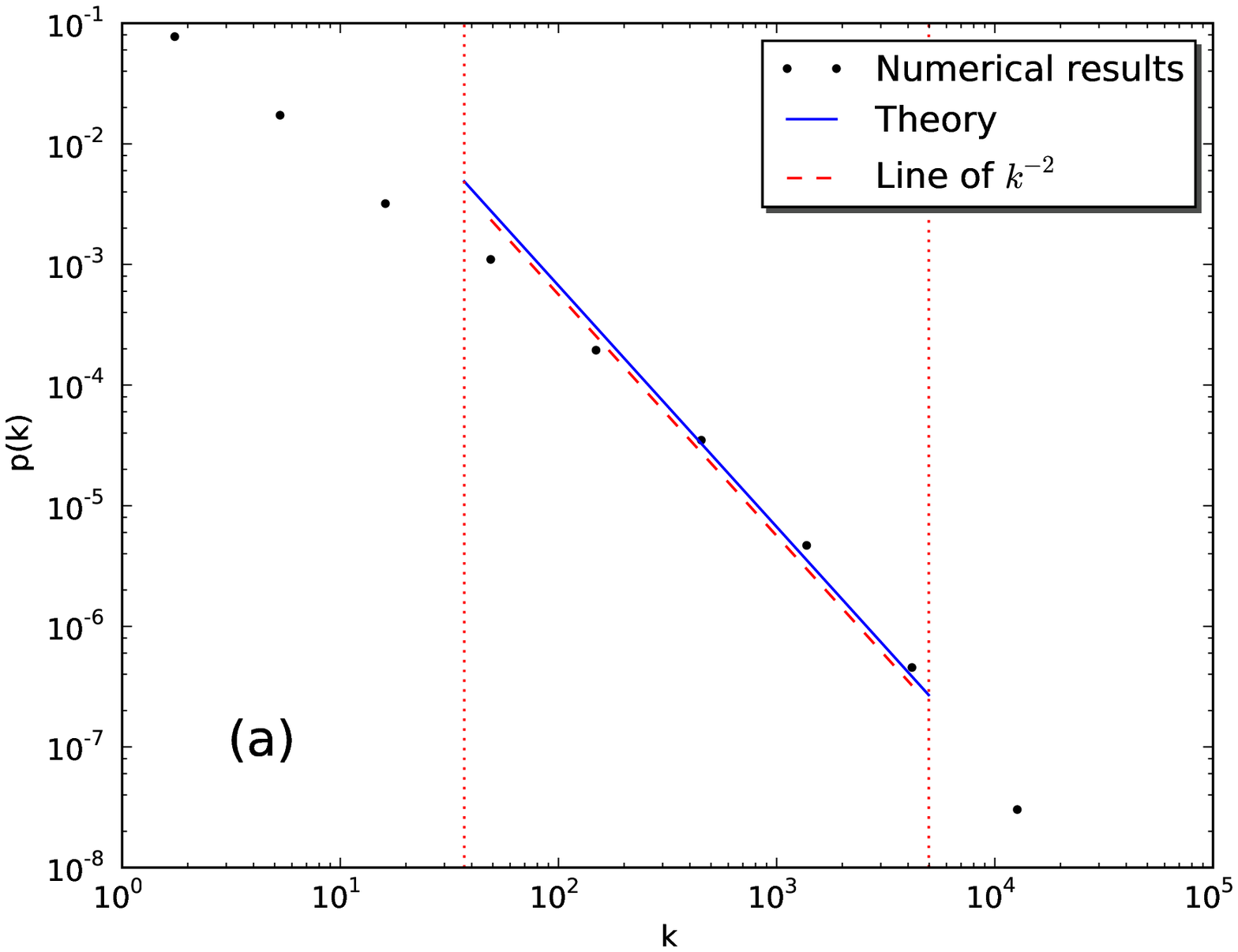}}
\resizebox{70mm}{!}{\includegraphics{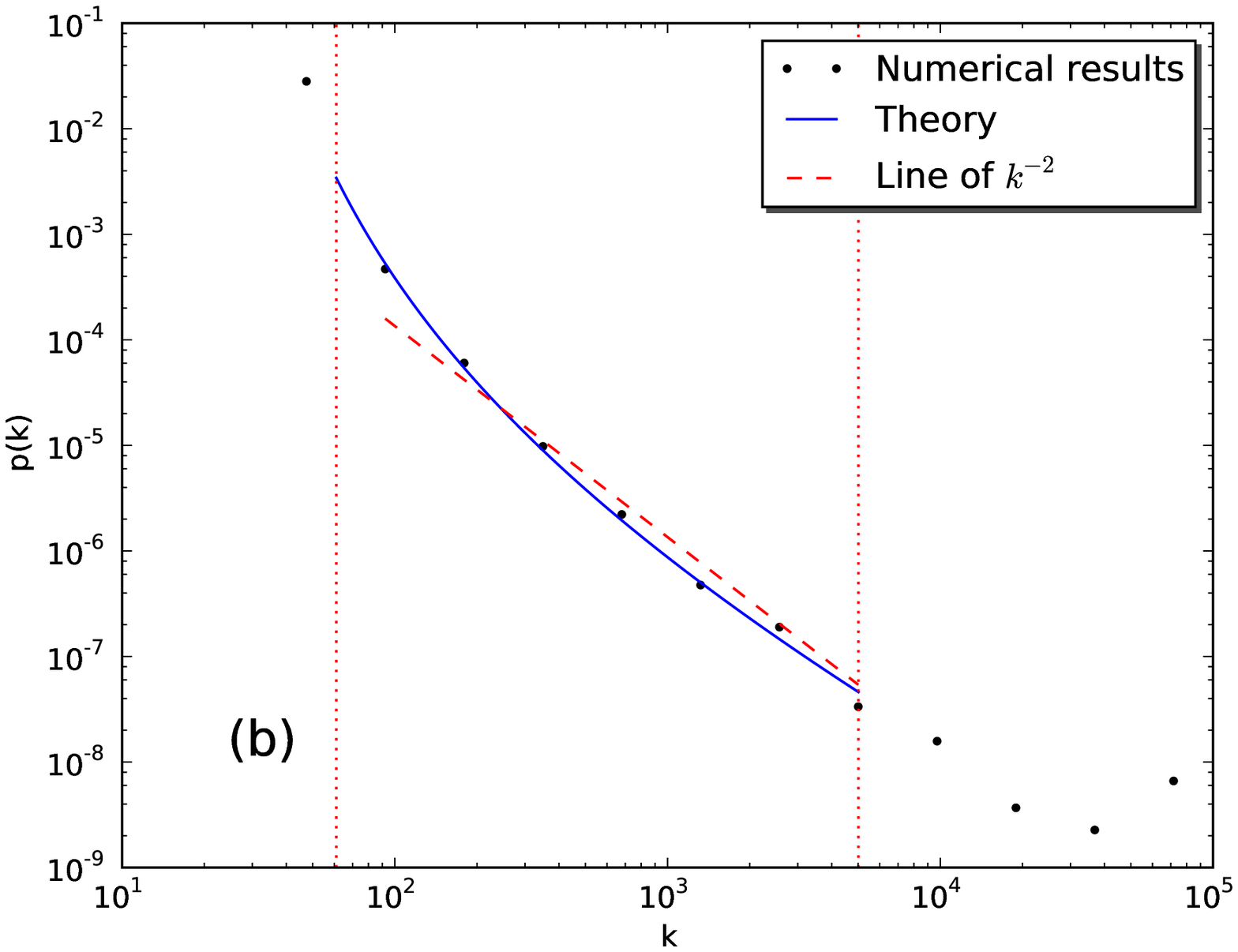}}
\resizebox{70mm}{!}{\includegraphics{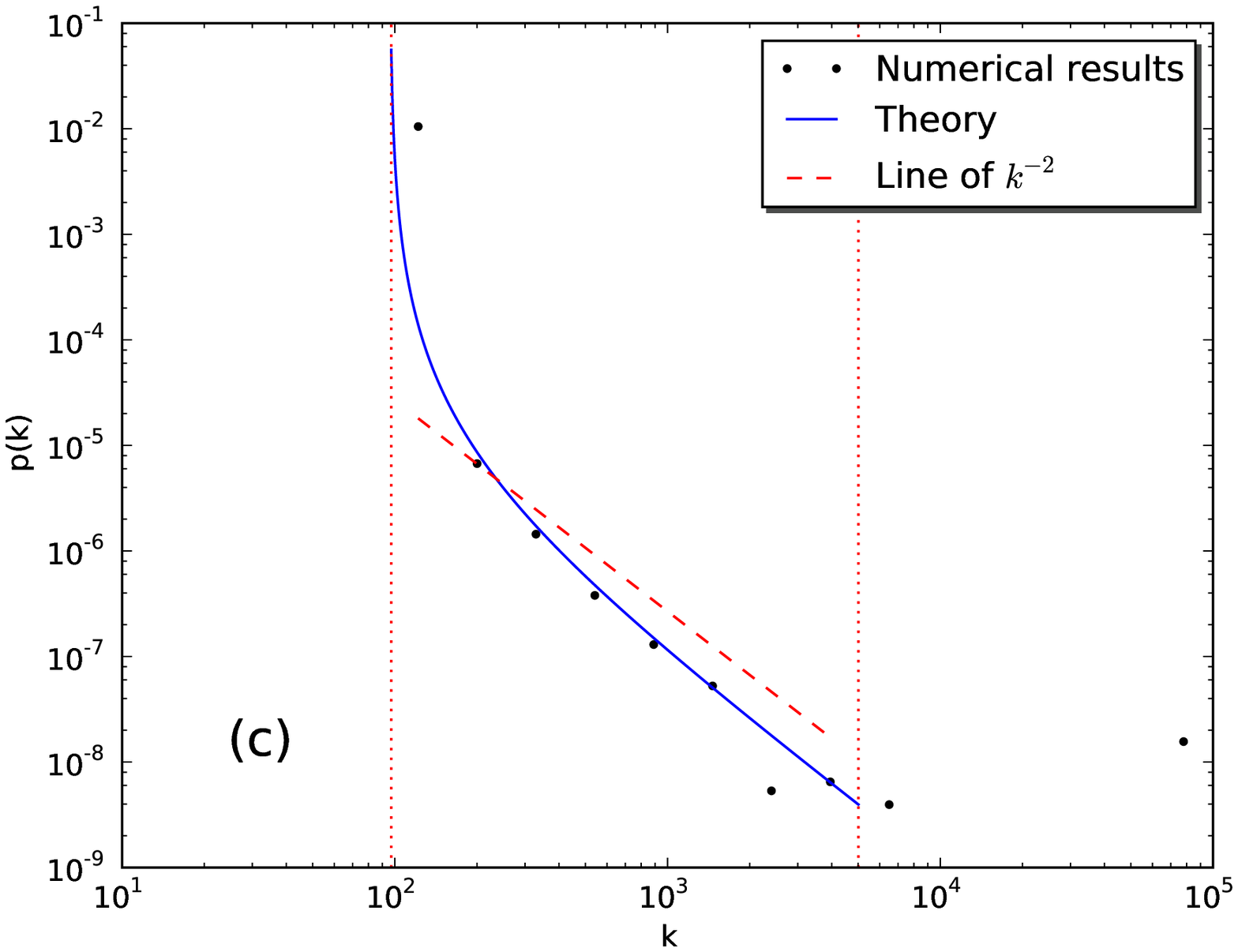}}
\resizebox{70mm}{!}{\includegraphics{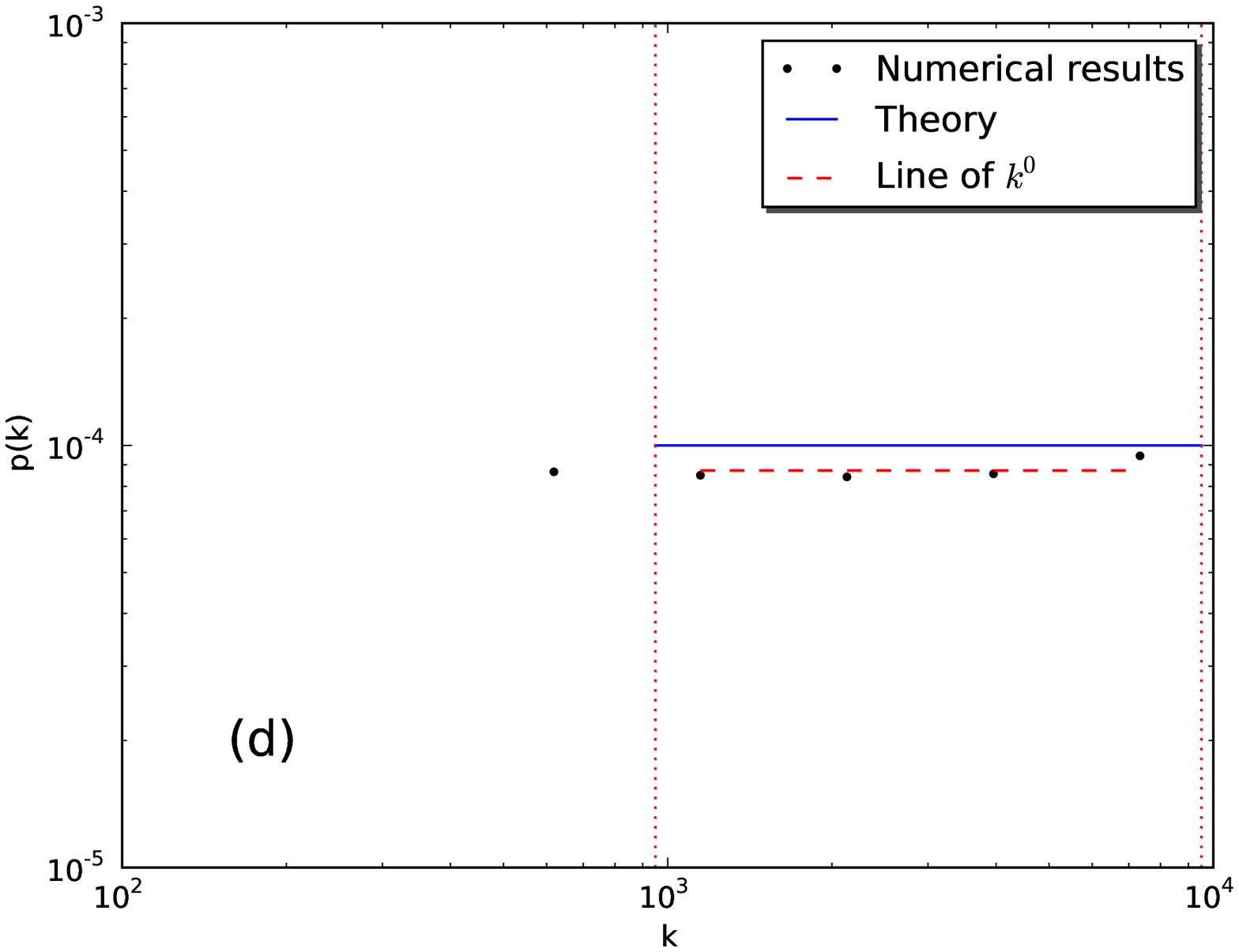}}
\caption{Numerical results for the degree distribution $p(k)$. 
(a) Weights are given by the normal distribution with $m = 0$, $s = 1$ 
in Table~\ref{tab:table1} ($\xi = 0$) and $n=10^5$, $\theta=5$, $u=1.65$, 
and $\tilde{\sigma}=0.47$. 
(b) Lognormal distribution with $M=0$, $S=1$ in Table~\ref{tab:table1} 
($\xi=0$) and $n=10^5$, $\theta=30$, $u=5.19$, and $\tilde{\sigma}=2.55$.
(c) Standard Fr\'echet distribution $F(w) = \exp( - 1 / x)$ ($\xi=1$)
and $n=10^5$, $\theta=1000$, $u=19.7$, and $\tilde{\sigma}=20.34$. 
(d) Uniform distribution in Table~\ref{tab:table1} ($\xi=-1$)
and $n=10^4$, $\theta=0.95$, $u=0.05$, and $\tilde{\sigma}=0.95$. 
 The two vertical dotted lines in each figure indicate the lower and upper 
bounds of the applicable range in Eq.~(\ref{eq:corr_range_k}). 
 The theoretical prediction 
in Eq.~(\ref{eq:p_k_ngtm_exact}) or (\ref{eq:pk_xi_zero}) is indicated by 
the solid curves, and power laws $k^{-s}$ with $s=0$ or $2$ are indicated 
by the dashed lines.}
\label{fig:degree_dist}
\end{figure*}
\begin{figure*}[tb]
\resizebox{70mm}{!}{\includegraphics{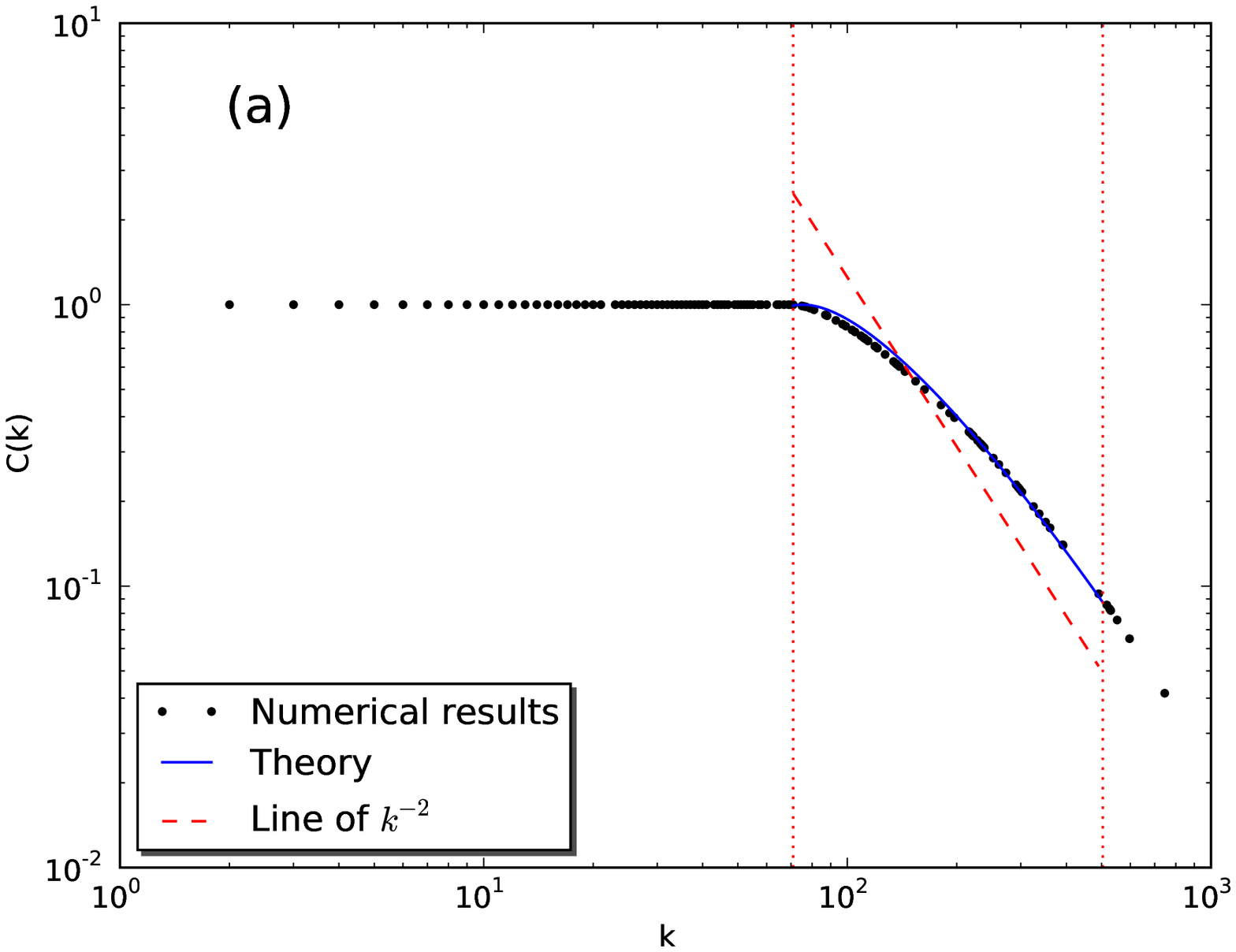}}
\resizebox{70mm}{!}{\includegraphics{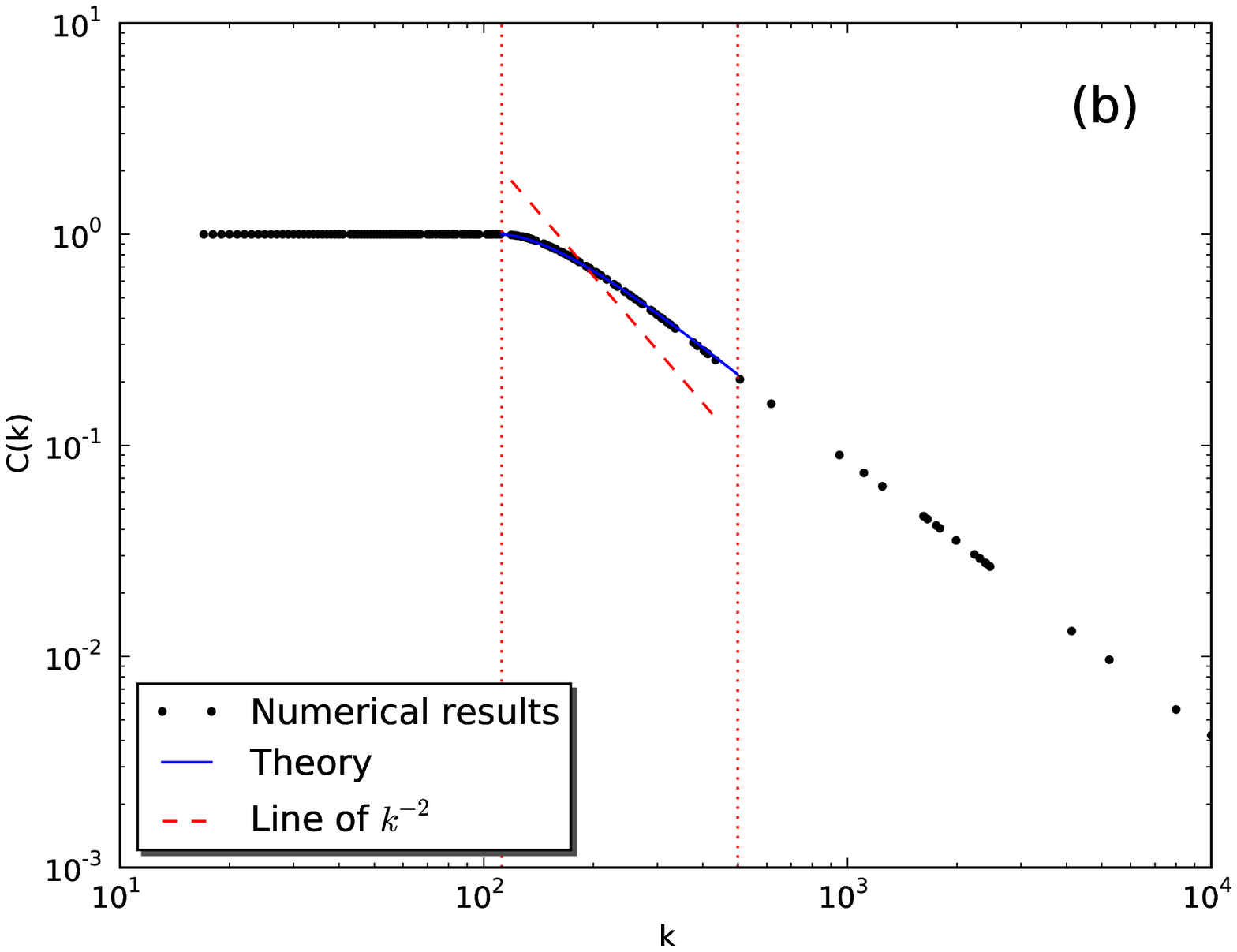}}
\resizebox{70mm}{!}{\includegraphics{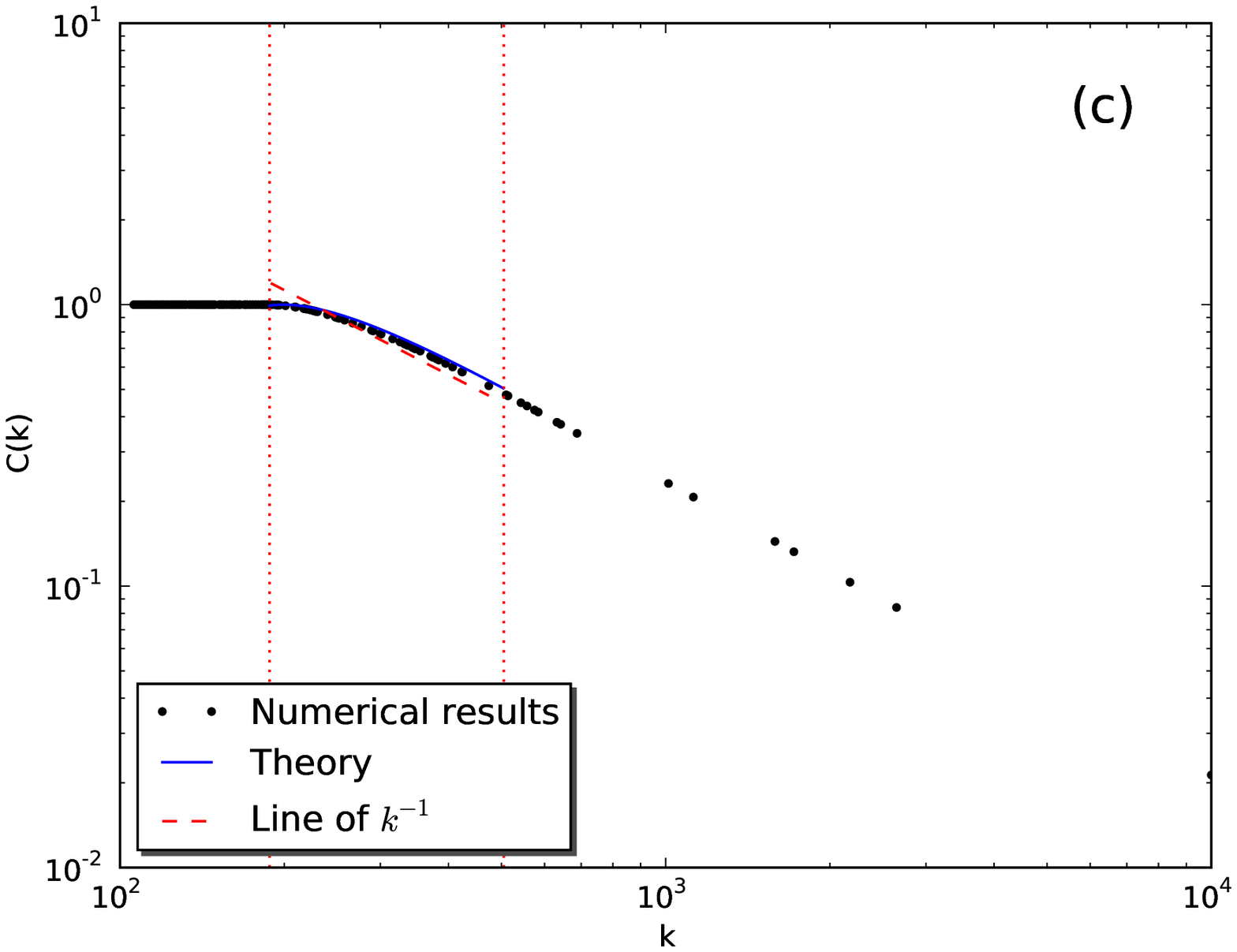}}
\resizebox{70mm}{!}{\includegraphics{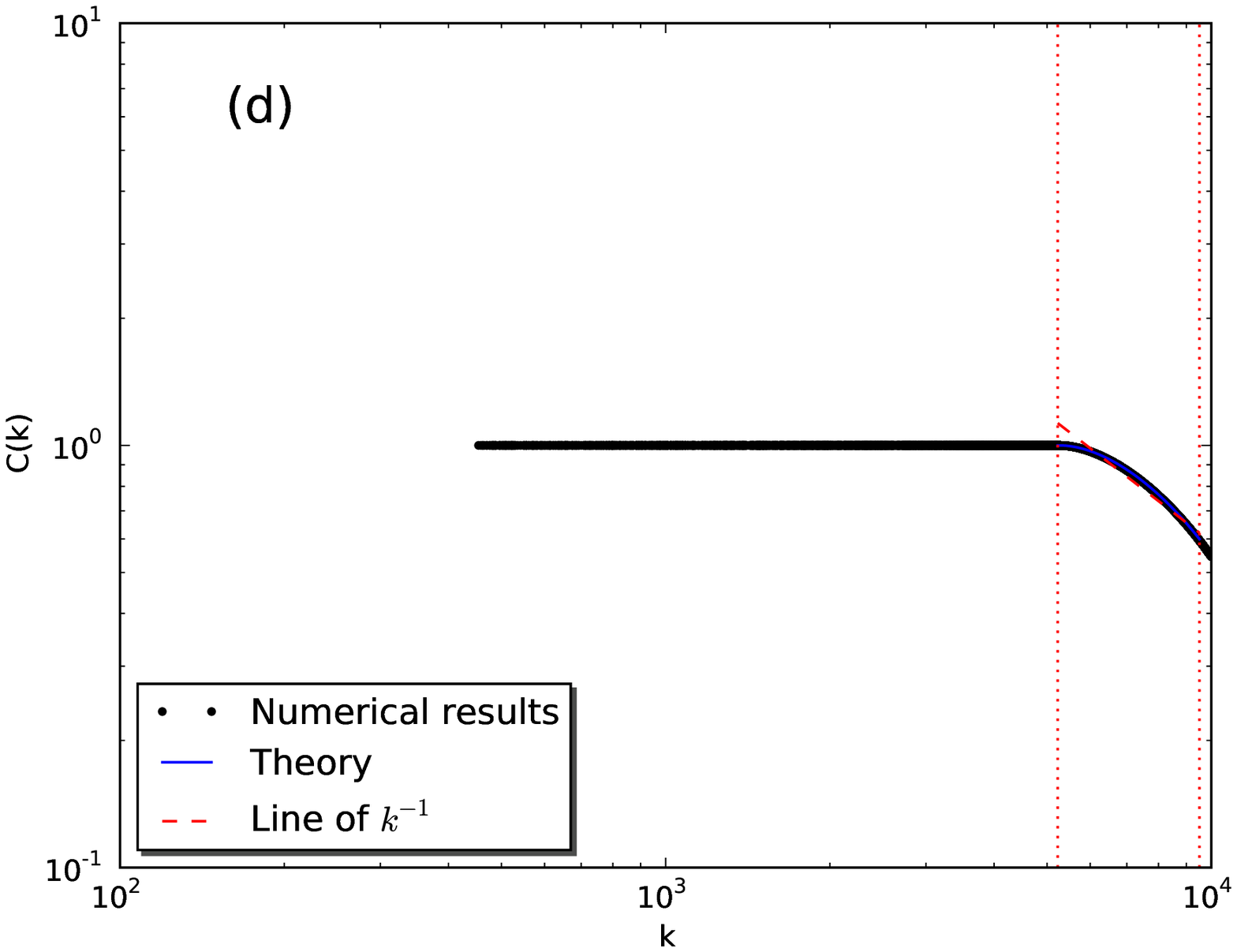}}
\caption{Numerical results for the degree-wise clustering coefficient $C(k)$. 
(a) Weights are given by normal distribution with $m = 0$, $s = 1$ 
in Table~\ref{tab:table1} ($\xi = 0$) and $n=10^4$, $\theta=4.9$, $u=1.64$, 
and $\tilde{\sigma}=0.50$. 
(b) Lognormal distribution with $M=0$, $S=1$ in Table~\ref{tab:table1}
($\xi=0$) and $n=10^4$, $\theta=20$, $u=5.18$, and $\tilde{\sigma}=2.69$.
(c) Standard Fr\'echet distribution $F(w) = \exp( - 1 / x)$ ($\xi=1$) 
and $n=10^4$, $\theta=100$, $u=19$, and $\tilde{\sigma}=20.20$. 
(d) Uniform distribution in Table~\ref{tab:table1} ($\xi=-1$) 
and $n=10^4$, $\theta=0.95$, $u=0.05$, and $\tilde{\sigma}=0.95$. 
 The two vertical dotted lines in each figure indicate the lower and upper 
bounds of the applicable range in Eq.~(\ref{eq:true_ck_range}). 
The theoretical prediction
in Eq.~(\ref{eq:ck_integral_form}) or (\ref{eq:Ck_xi_zero}) is indicated by 
the solid curves, and the power laws $k^{-s}$ with $s=1$ or $2$ are indicated
by the dashed lines.}
\label{fig:clustering_coeff_dist}
\end{figure*}
 This section presents numerical results for the distributions of degrees
and clustering coefficients to compare them with the theoretical results 
obtained in the previous section. 
 To observe power laws, we must select the values of parameters 
$n$, $\theta$, and $u$ optimally with the next two items in mind. 
 First, $\theta$ and $u$ are satisfied by the implicit condition, 
$\theta \ge 2 u$, that prevents the applicable ranges in 
Eqs.~(\ref{eq:corr_range_k}) and ~(\ref{eq:true_ck_range}) from 
diminishing. 
 Second, as the ranges increase as $n$ and $\theta - u$ increase,
we need to give them as numerically large values as possible 
to identify the theoretical curves in 
Eqs.~(\ref{eq:p_k_ngtm_exact}), (\ref{eq:pk_xi_zero}), 
(\ref{eq:ck_integral_form}), and (\ref{eq:Ck_xi_zero}) clearly in 
the corresponding ranges. 
 We determine the three parameters $n$, $\theta$, and $u$ using 
the following procedures.
 From the shape of the weight distribution function, appropriate $u$ and 
$\tilde{\sigma}$ are calculated using the fExtremes package \cite{Wuertz2007} 
in R language \cite{R2008}. When the value of $u$ is fixed, 
we give a large $\theta$ while retaining the condition $\theta \ge 2 u$ 
and gradually increase $n$ until the universal power laws can be 
observed widely as well. 
 Typical results where the weights are given by the normal and
the lognormal distributions ($\xi=0$), 
the standard Fr\'echet distribution ($\xi>0$),
and the uniform distribution ($\xi<0$) are given
in Figs.~\ref{fig:degree_dist} and ~\ref{fig:clustering_coeff_dist}. 
 As can be seen from these figures, the theoretically predicted curves 
nicely fit the numerical results, meaning 
Eqs.~(\ref{eq:p_k_ngtm_exact}), ~(\ref{eq:pk_xi_zero}), 
~(\ref{eq:ck_integral_form}), and ~(\ref{eq:Ck_xi_zero}) are all accurate.
 Each line of power law $k^{-s}$ with $s=0$, $1$, and $2$ in the figure 
indicates the universal power law in Eqs.~(\ref{eq:power-law1a}), 
~(\ref{eq:power-law1b}), ~(\ref{eq:powerlaw2a}), and ~(\ref{eq:powerlaw2b}).
 These lines are also extremely consistent with the numerical results.
 The plots of the numerical results around the lower bounds of the ranges
appear to deviate from being linear, i.e., the power laws,
because the original formulas for the
degree distribution in Eq.~(\ref{eq:p_k_ngtm_exact}) and 
the clustering coefficient in Eq.~(\ref{eq:ck_integral_form}) 
are only valid in large-degree regions.
 Therefore, all the deviations result from small-degree effects. 

 Consequently, we evaluated the theoretical predictions 
in Sec.~\ref{sec:main_result} and found they were in good agreement with 
the results obtained from numerical experiments 
in both the distributions of degrees and clustering coefficients. 

\section{Summary and Discussion}
\label{sec:SD}
 We analyzed the threshold network model based on extreme value 
theory.  Where the number of nodes, $n$, and threshold $\theta$ 
were sufficiently large, the weight distribution function, $F(w)$, 
could be assumed to belong to a family of the generic weight function in 
Eq.~(\ref{eq:evt_fx}) for $x > u$ as long as the limit distribution of 
maximum values for $F(w)$ converged to the distribution of GEVs in
Eq.~(\ref{eq:gev_fam_dist}). 
 As previously mentioned, many important weight functions are classified into 
this family; therefore, we could comprehensively treat the model 
with various weight functions.
 We then calculated the exact forms of the distributions of degrees and
clustering coefficients by applying the generic weight function in
Eq.~(\ref{eq:evt_fx}). 
 As a result, we found theoretical evidence for
the universal power laws in Eqs.~(\ref{eq:corr_range_k}) and 
~(\ref{eq:true_ck_range}), which was extremely consistent with the results 
obtained by Masuda \textit{et al.} \cite{MMK2004}. 
 We also demonstrated that even though the theoretical prediction assumed
a sufficiently large number of nodes, $n$, threshold value, $\theta$, 
and degree, $k$, the theoretical results could be reproduced 
through numerical experiments. This fact could explain that real complex 
networks with a large number of nodes and a high threshold (if they
exist) could universally appear with scale-free properties.

 We conclude by referring to recent mathematical studies 
on threshold network models \cite{KMRS2005,IKM2007,IKM2009}. 
 We think that our investigation was important as a prototype
to apply extreme value theory to various other models of complex networks.
 In future work, it would be interesting to reformulate our method 
mathematically to find the possibilities of extending its range of 
applications. 

\section*{Acknowledgments}
 The authors wish to thank Drs. Norio Konno, Yusuke Ide, and Naoki Masuda 
for the productive discussions we had with them and the useful comments
they provided.
 This work was partially supported by the Japan Society for the Promotion
of Science through Grants-in-Aid for Scientific Research (S) (18100001) 
and (C) (20500080).




\end{document}